\documentclass[10pt,journal,compsoc]{IEEEtran}
\usepackage{graphicx} 
\usepackage{subfigure}
\usepackage{utfsym}
\usepackage{listings}
\usepackage{algorithm}
\usepackage{algorithmicx}
\usepackage[noend]{algpseudocode}
\usepackage{amsmath}
\usepackage{array}
\usepackage{cleveref}
\usepackage{color}
\usepackage[most]{tcolorbox}
\usepackage{multirow}
\usepackage[normalem]{ulem}
\usepackage{booktabs}
\usepackage{pifont}
\usepackage{xspace}
\usepackage{enumitem}
\usepackage{threeparttable}
\usepackage{balance}
\usepackage[T1]{fontenc} 
\usepackage{amsmath,amsfonts,amssymb}
\usepackage{url}
\usepackage{cite}
\newcommand{\new}[1]{#1}

\newcommand{\name}{{FlexFL}\xspace}
\newcommand{\nameFL}{{Agent4SR}\xspace}
\newcommand{\nameRefine}{{Agent4LR}\xspace}

\begin{document}
\title{\name: Flexible and Effective Fault Localization with Open-Source Large Language Models}

\author{Chuyang Xu, 
        Zhongxin Liu, 
        Xiaoxue Ren,
        Gehao Zhang,
        Ming Liang,
        David Lo
\IEEEcompsocitemizethanks{\IEEEcompsocthanksitem Chuyang Xu, Zhongxin Liu and Xiaoxue Ren are with the State Key Laboratory of Blockchain and Data Security, Zhejiang University, Hangzhou, 310027, China. \\ 
E-mail: \{chuyangxu, liu\_zx, xxren\}@zju.edu.cn
\IEEEcompsocthanksitem Gehao Zhang and Ming Liang are with Ant Group, China. \\
Email: gehaozhang1999@gmail.com, liangming.liang@antgroup.com
\IEEEcompsocthanksitem David Lo is with the School of Computing and Information Systems, Singapore Management University, Singapore 188065 \\
E-mail: davidlo@smu.edu.sg
\IEEEcompsocthanksitem Zhongxin Liu is the corresponding author. 
}
}


\IEEEtitleabstractindextext{%
\begin{abstract}
Fault localization (FL) targets identifying bug locations within a software system, which can enhance debugging efficiency and improve software quality. 
Due to the impressive code comprehension ability of Large Language Models (LLMs), a few studies have proposed to leverage LLMs to locate bugs, i.e., LLM-based FL, and demonstrated promising performance.
However, first, these methods are limited in flexibility. They rely on bug-triggering test cases to perform FL and cannot make use of other available bug-related information, e.g., bug reports.
Second, they are built upon proprietary LLMs, which are, although powerful, confronted with risks in data privacy.
To address these limitations, we propose a novel LLM-based FL framework named \name, which can flexibly leverage different types of bug-related information and effectively work with open-source LLMs.
\name is composed of two stages.
In the first stage, \name reduces the search space of buggy code using state-of-the-art FL techniques of different families and provides a candidate list of bug-related methods.
In the second stage, \name leverages LLMs to delve deeper to double-check the code snippets of methods suggested by the first stage and refine fault localization results.
In each stage, \name constructs agents based on open-source LLMs, which share the same pipeline that does not postulate any type of bug-related information and can interact with function calls without the out-of-the-box capability.
Extensive experimental results on Defects4J demonstrate that \name outperforms the baselines and can work with different open-source LLMs.
Specifically, \name with a lightweight open-source LLM Llama3-8B can locate 42 and 63 more bugs than two state-of-the-art LLM-based FL approaches AutoFL and AgentFL that both use GPT-3.5. 
In addition, \name can localize 93 bugs that cannot be localized by non-LLM-based FL techniques at the top 1.
\new{Furthermore, to mitigate potential data contamination, we conduct experiments on a dataset which Llama3-8B has not seen before, and the evaluation results show that \name can also achieve good performance.}
\end{abstract}

}

\maketitle

\section{Introduction}
\label{section:introduction}
Fault localization (FL) is tasked with precisely identifying the locations of bugs within a software system~\cite{SurveyOfFL}.
Effective FL can significantly enhance the efficiency of debugging by concentrating on the buggy areas, improving software quality and developer productivity~\cite{FL_families}. 
To facilitate debugging, prior work has proposed many approaches to automatically localize buggy program entities (e.g., statements, methods, and files), among which information-retrieval-based techniques (IRFL) ~\cite{BLIA,FinerBench4BL,FineLocator,BoostNSift}, spectrum-based techniques (SBFL) ~\cite{Ochiai,Dstar,PRFL-MA,proFL} and hybrid fault localization (HybridFL)~\cite{SBIR,AML} have shown to be effective.

Recently, due to the capability of large language models (LLMs) in language understanding, planning, and reasoning~\cite{Codex,GPT-4,CodeLlama,llama3,Qwen2,Deepseek-coder,ReACT,ComparisonOfCodeLLMs,program_understanding}, LLMs have attracted significant attention from software debugging researchers with successful research done on patch generation~\cite{ConversationalAPR} and bug reproduction~\cite{LIBRO}.
These researches show that LLMs can effectively understand source code and bug-related information (e.g., failed tests), plan the debugging process, and reason the root causes of bugs, and thus can be beneficial for FL.
A few studies~\cite{LLMinFL, AutoFL, AgentFL,reason4explainableFL} proposed to automate FL with LLMs.
Wu et al.~\cite{LLMinFL} leverage GPT-3.5 and GPT-4 to localize buggy statements in the context of their belonging method or class. 
However, it is difficult to use as a standalone FL technique in practice because locating faulty methods or classes remains a difficult problem for existing FL techniques.
AgentFL~\cite{AgentFL} and AutoFL~\cite{AutoFL} focus on localizing the buggy methods from an entire project, which are more practical and achieve promising results.
AgentFL~\cite{AgentFL} starts with bug-triggering test cases (for short, trigger tests) and leverages a manually crafted and fixed procedure to iteratively prompt ChatGPT to identify buggy locations.
Kang et al.~\cite{AutoFL} proposed AutoFL that equips GPT3.5 and GPT-4 with function calls (i.e., external tools or programs provided by users) to get the classes and methods covered by trigger tests and access the actual code snippets in the project, achieving state-of-the-art performance.



LLM-based FL techniques~\cite{AgentFL,AutoFL} have outperformed traditional non-LLM-based FL techniques.
However, they have the following limitations.
(1) Their flexibility in handling different types of bug-related information is limited. 
AgentFL and AutoFL both rely on trigger tests to build pipelines or agents.
Therefore, them can only work when trigger tests are available and cannot leverage other bug-related information, e.g., bug reports.
However, in practice, the available bug-related information for different bugs may differ.
For example, only bug-triggering test cases are available for the bugs found by fuzzing tools, while the bugs reported by users may only have bug reports for FL.
Furthermore, leveraging all available bug-related information can improve FL performance~\cite{SBIR,AML}.
(2) They are based on closed-source LLMs, which are confronted with concerns about data privacy. 
Existing LLM-based FL techniques~\cite{AgentFL,AutoFL,LLMinFL} all leverage GPT-3.5 or even GPT-4, which have demonstrated powerful capacities for instruction compliance and task solving.
It is still unknown whether they can work with open-source LLMs, which are widely used by organizations concerned with data privacy but have limited context length and inferior performance.
In addition, AutoFL ~\cite{AutoFL} leverages the out-of-the-box function calling capability of OpenAI GPT\footnote{https://platform.openai.com/docs/guides/gpt/function-calling}.
This capability enables users to merely provide function descriptions for the GPT, which can accurately respond with a complete function call with arguments in the JSON format. 
To the best of our knowledge, this capability is not provided by most open-source LLMs.

To fill these gaps, we propose a novel LLM-based FL framework \name.
\name comprises two stages: \ding{172} space reduction and \ding{173} localization refinement.
In its first stage, FlexFL utilizes an LLM-based agent named \nameFL and non-LLM-based FL approaches to generate a candidate list of suspicious methods.
In the second stage, \name constructs another agent named \nameRefine, which can fully leverage LLMs’ capabilities for code comprehension and reasoning to focus on checking the code snippets of the recommended methods in the candidate list.

\new{\name differentiates itself from existing LLM-based FL techniques, i.e., AutoFL~\cite{AutoFL} and AgentFL~\cite{AgentFL}, as follows: 
\textbf{(1) Unique framework:} The two-stage process of \name allows it to integrate existing LLM-based FL techniques for space reduction and thus can be complementary to them (shown in Section~\ref{section:RQ1.1}).
The design of the space reduction stage provides LLMs with information extracted using diverse strategies to refer to, such as text similarity from IRFL techniques and dynamic coverage information from SBFL approaches. 
These FL techniques complement each other and enhance the overall effectiveness of FL (demonstrated in Section ~\ref{section:ablation_study}), while AutoFL and AgentFL only rely on LLMs.
The design of the localization refinement stage mitigates the impact of the limited context length of LLMs and their inferior performance on long input contexts~\cite{liu2024lost}, which is more severe for open-source LLMs, and thus enables effective fault localization with open-source LLMs. 
\textbf{(2) Enhanced flexibility:} FlexFL designs its framework and LLM-based agents without assuming the existence of any specific type of bug-related information, which enables it to leverage flexible types of bug-related information, e.g., trigger tests and bug reports, and can significantly improve the performance (shown in Section~\ref{section:input_ablation}). 
In contrast, AutoFL and AgentFL both rely on trigger tests to build their agents or frameworks and cannot utilize other types of bug-related information, such as bug reports. 
\textbf{(3) Strong adaptation:} FlexFL designs a pipeline for LLMs to interact with function calls via a Reason-Act Framework~\cite{ReACT} and a postprocessing process, which can assist various chat models in code exploration of software repositories, including lightweight open-source LLMs (shown in Section~\ref{section:RQ1.1} and Section~\ref{section:RQ3}). 
However, AgentFL lacks such a pipeline, and AutoFL relies on the out-of-the-box function calling capability of OpenAI GPT, which is not provided by most open-source LLMs, to build its agent pipeline.
These limit their adaptation.
}

We perform thorough experiments to evaluate the effectiveness of \name and the contributions of its components.
We evaluate \name on a widely used debugging benchmark Defects4J~\cite{Defects4J}.
Experimental results show that \name outperforms non-LLM-based FL baselines and existing LLM-based techniques that use GPT-3.5 in all metrics. 
For example, \name with Llama3-8B-Instruct~\cite{llama3}, a lightweight open-source LLM, can localize 12.3\%, 18.9\%, 21.6\% more bugs than the state-of-the-art LLM-based approach AutoFL with GPT-3.5, in top-1, top-3, and top-5, respectively.
We also implement \name based on two other lightweight open-source LLMs, i.e., Qwen2-7B-Instruct~\cite{Qwen2} and Mistral-Nemo-12B-Instruct~\cite{Mistral-Nemo}, of which the comparable performance indicates \name is generalizable across different lightweight open-source LLMs.
To investigate the impact of potential data leakage of LLMs, 
we evaluate \name on 28 recently fixed bugs from the GHRB~\cite{GHRB} dataset, where \name also achieves good performance.
These experimental results show that \name achieves flexible and effective fault localization based on open-source LLMs.

In summary, we make the following contributions.
\begin{itemize}
\item We propose a flexible and effective fault localization framework named \name, which can handle different types of bug-related information and is effective.
\item We enable the construction of agents based on open-source LLMs for FL. To the best of our knowledge, this is the first attempt to build FL agents based on open-source LLMs.
\item We comprehensively evaluate \name, and evaluation results show that our framework outperforms the baselines by substantial margins.
\item We open source our replication package~\cite{package}, including the dataset, the source code of \name, and experimental results, for follow-up works.
\end{itemize}

The remainder of the paper is organized as follows.
Section ~\ref{section:related_works} gives an overview of the relevant literature, and Section~\ref{section:approach} describes our approach.
We illustrate evaluation settings and research questions in Section ~\ref{section:setting}.
After elaborating on our evaluation results in Section ~\ref{section:evaluation}, we analyze Agent4SR's failure cases, compare \name with learning-based FL techniques, and discuss threats to validity in Section ~\ref{section:discussion}. 
Section ~\ref{section:conclusion} concludes and describes future work.

\section{Background and Related Work}
\label{section:related_works}
Our framework \name focuses on method-level FL, i.e., locating buggy methods from an entire project. 
In this section, we review existing non-LLM-based FL techniques and then introduce the related LLM-based FL techniques.

\subsection{Non-LLM-based Fault Localization}
Typical method-level FL techniques include Spectrum-based Fault Localization (SBFL), Information Retrieval-based Fault Localization (IRFL), Mutation-based Fault Localization (MBFL), and Hybrid Fault Localization (HybridFL), which combines two or more FL techniques.

Information Retrieval-based Fault Localization (IRFL)~\cite{IRFL} measures the textual similarity between bug reports and program entities, and outputs a ranked list of program entities as suspicious bug locations. 
To the best of our knowledge, only a few IRFL techniques can localize bugs at the method level~\cite{FinerBench4BL, FineLocator, BLIA1.5, BoostNSift}.
Among them, BoostNSift~\cite{BoostNSift} achieves the state-of-the-art method-level FL performance by embedding query boosting and code sifting in conjunction with the BM25 Information Retrieval (IR) model.
In the query boosting step, BoostNSift adds weights to the title field of the bug report. 
In code sifting, the relevance of a program entity to a specific bug report is compared against its relevance to a collection of bug reports.

Spectrum-based Fault Localization (SBFL) is one of the most prevalent fault localization techniques~\cite{FL_families}, which are broadly adopted in program debugging~\cite{Ochiai,SBIR}. 
SBFL techniques analyze the run-time behavior of the passing and failing test cases and rank program entities based on program spectrum~\cite{FL_families}. 
Traditional formula-based SBFL techniques calculate suspiciousness scores using a ranking metric, e.g., Ochiai~\cite{Ochiai}, Dstar~\cite{Dstar}, and Tarantula~\cite{Tarantula}.

Mutation-based Fault Localization (MBFL)~\cite{MBFL} injects changes to each program entity (based on mutation testing~\cite{mutaion_testing}) to check its impact on the test outcomes.
Different from SBFL techniques, which consider whether a statement is executed or not, MBFL techniques~\cite{Muse, Metallaxis, wang2023contribution} consider whether the execution of a statement affects the result of a test by injecting mutants. 
The more often a statement affects failing tests, and the less often it affects passing tests, the more suspicious the statement is considered.
For a statement $s$, an MBFL technique generates a set of mutants $mut(s)$, assigns each mutant a score $M (m)$, and aggregates the $M (m)$ to yield a statement suspiciousness score $S(s)$.
MUSE~\cite{Muse} and Metallaxis-FL~\cite{Metallaxis} are two state-of-the-art MBFL techniques, both of which use different formulas to calculate $M (m)$ and different strategies for aggregation.

Hybrid Fault Localization (HybridFL)~\cite{FL_families,SBIR} combines results of different FL techniques.  
CombineFL~\cite{FL_families}, DeepFL~\cite{DeepFL}, and FLUCCS~\cite{FLUCCS} use learning-to-rank~\cite{learning-to-rank} machine learning approaches such as RankSVM~\cite{RankSVM} to combine multiple FL techniques, which need additional datasets for training.
SBIR~\cite{SBIR} uses the Monte Carlo rank aggregation algorithm~\cite{MonteCarlo} to combine IRFL and SBFL techniques’ ranked lists, which is an unsupervised HybridFL approach and achieves state-of-the-art performance.

\subsection{LLM-based Fault Localization}
Large language models (LLMs) have shown remarkable effectiveness in solving complex software engineering problems \cite{LIBRO, vaithilingam2022expectation, wang2024software, zhang2022repairing}. 
This success has drawn some attention from the fault localization research community~\cite{LLMinFL,AutoFL,AgentFL,reason4explainableFL}.
Specifically, Wu et al. \cite{LLMinFL} have assessed the effectiveness using GPT-3.5 and GPT-4 in locating buggy statements from a given code snippet, e.g., a buggy class or a buggy method.
Different from Wu et al.'s work, our work focuses on method-level FL and does not assume a known buggy code snippet.
AutoFL~\cite{AutoFL} starts from the classes and methods covered by failing test cases and leverages the function-calling capability of GPT-3.5 and GPT-4 to inspect code and pinpoint buggy methods from a project.
AgentFL~\cite{AgentFL} takes failing test cases as input and prompts ChatGPT multiple times with diversified information to handle manually designed tasks in each step of its process.
\new{
Unlike AutoFL and AgentFL, FlexFL is a novel and effective two-stage framework that leverages LLMs to refine localization results obtained by one or more FL approaches. 
As discussed in the introduction, FlexFL does not limit its input to a specific type of bug-related information and is more flexible than existing LLM-based FL approaches. 
In addition, FlexFL assists LLMs in repository-level code exploration via a Reason-Act Framework and a postprocessing process and utilizes existing FL techniques to help reduce search space, allowing effective fault localization with various chat models. 
In contrast, it remains unknown whether OpenAI-GPT-based AgentFL and AutoFL can effectively work with open-source LLMs. 
Moreover, the first stage of FlexFL can also use existing LLM-based FL approaches (shown in Section~\ref{section:RQ1.1}). 
To this end, FlexFL is also complementary to them.  
}

\section{Methodology}
\label{section:approach}
We propose \name, a flexible and effective LLM-powered framework for method-level fault localization. This section first introduces the overall framework of \name, then describes the design of two agents in \name (i.e., \nameFL and \nameRefine).

\begin{figure}[t]
\centering
\includegraphics[width=0.45\textwidth]{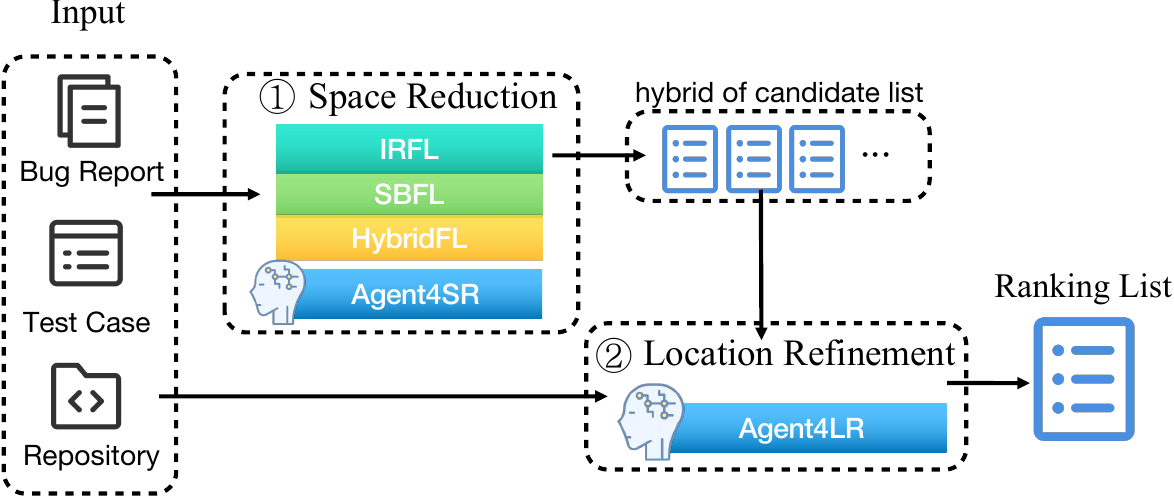}
\caption{The Overall Framework of \name}
\label{fig:framework}
\end{figure}

\subsection{Overview of \name}
Figure~\ref{fig:framework} shows the overall framework of our \name. 
Given the repository containing the bug and available bug-related information, e.g., the bug report or/and test cases, \name produces a ranking list of the top k most suspicious methods responsible for the bug.
Specifically, \name contains two stages \textbf{\ding{172} space reduction} and \textbf{\ding{173} localization refinement}.
In the first stage, \name utilizes an LLM-based agent named \nameFL and non-LLM-based FL approaches, e.g., IRFL, SBFL, and HybridFL techniques, to obtain a hybrid of candidate list that contains m suspicious methods.
In the second stage, \name leverages another LLM-based agent named \nameRefine to double-check the code snippets of the suggested methods in the candidate list and localize the top-k most suspicious methods.
The design of the two-stage process can help fully leverage LLMs' powerful capacity in language understanding and reasoning to focus on the most suspicious methods, enabling effective fault localization.

\subsubsection{Input of \name}\label{section:input}

\begin{table*}[htbp]
    \centering
    \caption{Text Description of Input (Example Bug: Time-25 in Defects4J~\cite{Defects4J})}
    \label{tab:input}
    \begin{tabular}{l|l}
        \toprule
        \multirow{4}{*}{\textbf{Bug Report}} & \textbf{\textit{Title: }}\#90 DateTimeZone.getOffsetFromLocal error during DST transition \\
        & {\textbf{\textit{Description: }}}This may be a failure of my understanding, but the comments in DateTimeZone.getOffsetFromLocal lead me to \\
        & believe that if an ambiguous local time is given, the offset corresponding to the later of the two possible UTC instants will be\\
        & returned - i.e. the greater offset\ldots\ldots(More details in~\cite{Time-25})\\
        \midrule
        \multirow{7}{*}{\textbf{Trigger Test}} &  public void test\_DateTime\_constructor\_Moscow\_Autumn() \{\\
        & DateTime dt = new DateTime(2007, 10, 28, 2, 30, ZONE\_MOSCOW);\\     
        & assertEquals("2007-10-28T02:30:00.000+04:00", dt.toString());\\
        & {\textbf{\textit{The last line shown above failed with the following stack trace.}}}\\
        & junit.framework.ComparisonFailure: expected:<...10-28T02:30:00.000+0[4]:00> but was:<...10-28T02:30:00.000+0[3]:00>\\
        & \sout{at junit.framework.Assert.assertEquals(Assert.java:100)}\\
        & at org.joda.time.TestDateTimeZoneCutover.test\_DateTime\_constructor\_Moscow\_Autumn(TestDateTimeZoneCutover.java:922)\\
        \bottomrule
    \end{tabular}
\end{table*}
Our \name incorporates two flexible types of bug-related information, i.e., bug reports and test suites, as input, while existing LLM-based FL techniques~\cite{AutoFL,AgentFL} can only leverage bug-triggering test cases (for short, trigger tests) in test suites.
Following these studies, we input trigger tests to LLMs in text, which contains the test methods and their stack traces.
As shown in Table~\ref{tab:input}, we remove the methods that do not belong to the buggy program (e.g., \texttt{at junit.framework.Assert.
assertEquals(Assert.java:100)}) from the stack trace, and truncate the test method at the point of failure to highlight critical information and save context length.
For bug reports, which are neglected by previous LLM-based works, we construct the text descriptions in the format shown in Table~\ref{tab:input}, which explicitly points out the titles and descriptions.
Note that these inputs (i.e., bug reports and trigger tests) can be processed not only individually but also in combination by LLMs via a dynamic prompt, offering \name the flexibility in addressing various fault localization scenarios.
For non-LLM-based FL approaches, these two types of bug-related information are input in the suitable ways they require.
Specifically, bug reports are also used in their text form for lexical match in IRFL techniques, and test suites are executed for collecting dynamic execution information (e.g., program spectrum) in SBFL techniques.

A critical input for FL tasks is the entire software repository containing the bug.
However, it is costly and ineffective to directly feed the whole large program into LLMs for processing since the computational complexity of existing LLMs is quadratic and the performance of LLMs degrades as the length of input contexts increases~\cite{liu2024lost}.
To solve this problem, the prior work AutoFL~\cite{AutoFL} allows LLMs to navigate the source code by calling custom-designed functions that return the information of the classes and methods covered by trigger tests, and access the implementation and documentation of any covered method. 
Inspired by AutoFL, we also designed a set of custom-designed function calls for LLMs to enable code exploration and relevant information extraction from the buggy program.
However, different from AutoFL where the designed function calls require coverage information collected with trigger tests, our designed function calls do not rely on any type of bug-related information and thus ensure the flexibility of \name.

\subsubsection{Space Reduction}\label{section:SR}
This stage aims to effectively narrow down the search space before localizing the buggy methods based on the constructed input.
Previous studies~\cite{AutoFL,AgentFL} have shown that LLM-based agents can automatically search for bug-related methods over a large software repository, which can be beneficial for reducing the search space.
Therefore, we design an agent named \nameFL based on open-source LLMs, which aims to reduce the bug-related code space via global searching in the buggy program.
In addition, existing non-LLM-based FL techniques have been proven helpful and valuable for filtering out unrelated methods in the buggy program~\cite{Ochiai,BoostNSift,SBIR,Muse,FL_families}.
Based on this idea, we propose to combine \nameFL with existing non-LLM-based FL techniques to complement \nameFL and better reduce the search space.
Specifically, given input introduced in Section~\ref{section:input}, \nameFL and non-LLM-based FL techniques respectively localize the top-k most suspicious methods responsible for the bug.
To combine the suggested methods of \nameFL and non-LLM-based FL techniques, we first place the results of \nameFL (i.e., top-k buggy methods) at the end based on the assumption that the methods localized by \nameFL are more likely to be localized by \nameRefine, which is also an LLM-based agent, so we do not need to emphasize them via high ranking.
Then, the remaining $m - k$ methods are divided equally among the non-LLM-based FL techniques and the top-k results of one technique would be followed by the top-k results of another.
The order of non-LLM-based FL techniques is based on another assumption that more precise localization results should be assigned a higher ranking to assist LLMs in refinement.
Finally, after space reduction, we can obtain a relatively comprehensive candidate list that contains $m$ suspicious methods.
These methods are localized by different techniques in various ways, thus containing diverse kinds of information beneficial for localization refinement.
For instance, \nameFL focuses on the semantic information extracted from the textual description of trigger tests while SBFL approaches emphasize dynamic execution information of test suites.

In this stage, \name employs existing non-LLM-based FL techniques and thus is enabled to process any type of input that previous approaches can handle.

\subsubsection{Localization Refinement}
For localization refinement, we further design an agent, named \nameRefine, which localizes the top-k most suspicious methods based on the textual description of bug-related information and the candidate list produced in the space reduction stage.
Unlike \nameFL, \nameRefine is aimed at double-checking the suspicious methods in the candidate list.
This design enables \nameRefine to use more tokens for planning, reasoning, and understanding, instead of code exploration, thus alleviating LLMs' limitation in processing long context.
\new{
Theoretically, \nameRefine has a chance to localize the buggy method as long as the buggy method is included in the candidate list produced by one of the FL techniques used in the space reduction stage.
}
More details can be found in Section~\ref{sections:Agent4LR}.

\subsection{Design of Agents}\label{section:agent}
The LLM-based agents within our \name, i.e., \nameFL and \nameRefine, are constructed following the same pipeline based on open-source LLMs.
Below, we first outline the overall pipeline of these agents and the function calls designed to assist them. 
Then, we respectively introduce the detailed designs of \nameFL and \nameRefine.

\begin{figure*}[htbp]
\centering
\includegraphics[width=1.0\textwidth]{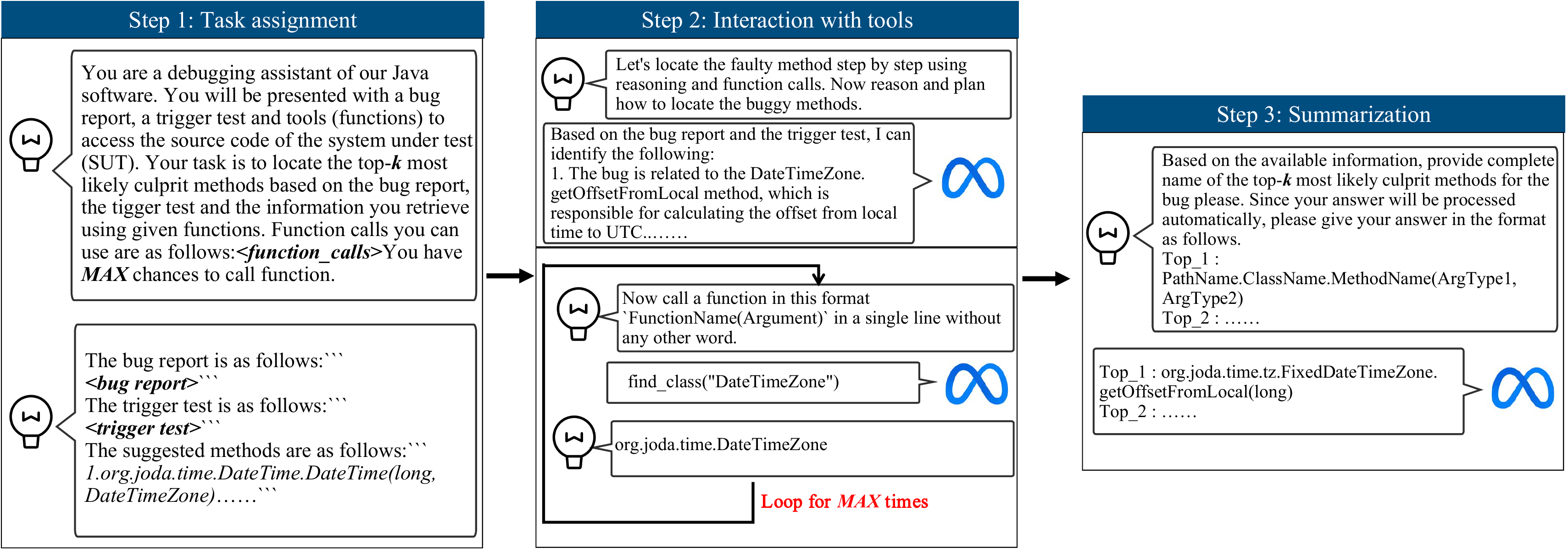}
\caption{The pipeline of agents. Bold text in <> indicates placeholders for input contents or description of function calls designed in Section~\ref{section:function_call}}
\label{fig:agent_pipeline}
\end{figure*}

\subsubsection{Pipeline of agents}\label{section:agent_pipeline} 
To prompt open-source LLMs to localize bugs with bug-related information, both agents in \name are designed to follow a three-step process: task assignment, interaction with function calls, and summarization. 
Figure~\ref{fig:agent_pipeline} uses the bug Time-25~\cite{Time-25} from Table~\ref{tab:input} as an example to illustrate this pipeline in detail.

\noindent{\textbf{Step 1: Task assignment.}} 
In this step, we make agents in \name flexibly handle bug reports and test cases both separately and together via the dynamic prompt.
Specifically, we design two prompts containing bug-related information and descriptions of available function calls.
The first prompt is a system prompt, as shown in Figure~\ref{fig:agent_pipeline}, which guides LLMs in their roles as debugging assistants.
The task assigned to LLMs is to localize the top k most suspicious methods based on the available bug-related information and the information extracted from the buggy programs via function calls (see Section~\ref{section:function_call}).
Note that if any type of bug-related information is not available, contents relative to it will be deleted from this prompt.
For instance, when only bug reports are available, we will instruct the agent to \textit{localize the top-k most suspicious methods based on the bug report and the information you retrieve} in the system prompt that is different from that shown in Figure~\ref{fig:agent_pipeline}.
The second prompt presents the descriptions of available inputs to the agent.
For both agents, text descriptions of bug reports and trigger tests (see Table~\ref{tab:input}) are given in order if available.
The candidate list of suggested methods that is an additional type of input for \nameRefine will be placed after descriptions of bug-related information. 

\noindent{\textbf{Step 2: Interaction with function calls.}} 
In this step, we design a pipeline for LLMs to interact with function calls via prompt engineering and a postprocessing process, which can assist any chat model in code exploration of software repositories, including lightweight open-source LLMs that do not support the out-of-the-box function calling capability.

Following the insights from ReAct~\cite{ReACT}, we first prompt LLMs to \textit{reason and plan how to use function calls} to help LLMs organize their thoughts and develop a strategy for localizing buggy methods, as shown in Figure~\ref{fig:agent_pipeline}.
Through reasoning, the LLMs can form a clear hypothesis about where the bug might be and why it is happening. 
This prepares them for interaction with function calls, where they can act on these hypotheses to gather more specific information or confirm their hypotheses.
For example, the reasoning results of Time-25 show its root cause, i.e., \textit{The bug is related to the DateTimeZone.getOffsetFromLocal method, which is responsible for calculating the offset from local time to UTC}.

After reasoning, LLMs are prompted to interact with function calls that assist in extracting detailed information from the buggy programs.
Different from proprietary LLMs like OpenAI GPT, open-source LLMs do not have the capability of function calling and cannot accurately identify the function calls needed to be called and precisely construct their arguments.
To enable function calling based on open-source LLMs for FL, we first design a prompt that asks LLMs to \textit{call a function in the format `FunctionName(Argument)` in a single line without any other word}.
Then we process the function call provided by LLMs, e.g., \texttt{find\_class("DateTimeZone")} shown in Figure~\ref{fig:agent_pipeline}, via the regular expressions that are consistent with the required format.
After obtaining the name (e.g. \texttt{find\_class}) and arguments (e.g. \texttt{"DateTimeZone"}) of the function call, we match the name and call the corresponding function with the arguments to extract information from the buggy program.
Finally, we append the extracted information (e.g., \texttt{org.joda.time.DateTimeZone} in Table~\ref{tab:input}) to the context and enter the next loop.
If the name of the function call provided by LLMs is not included in the given set, we return a prompt \textit{Please call functions in the right format `FunctionName(Argument).`}.

Such a conversation will loop \textit{MAX} times, which is also specified in the system prompt.
If the whole conversation exceeds the maximum context length of the used LLM, we decrease the value of \textit{MAX} by 1 and rerun this pipeline.
In addition, to conserve time and computational resources and further prevent exceeding the context length limit of LLM, we design a fixed and special function call \texttt{exit()}, which does not return information and is provided for LLMs to exit the step of interaction with function calls when LLMs are confident to localize buggy methods.
Therefore, this iterative process of information gathering and analysis continues until either the LLMs have performed \textit{MAX} function calls or issued an \texttt{exit} command.
This structured interaction ensures that the LLMs gather comprehensive information necessary for precise bug identification.

\noindent{\textbf{Step 3: Summarization.}} 
In this step, we instruct LLMs to summarize and pinpoint the top k most suspicious methods via integrating inputs from Step 1 and the information retrieved in Step 2. 
To ensure full automation and standardization, the responses from the LLMs are formatted according to a predetermined schema, i.e., \textit{Top\_i: PathName.ClassName.MethodName( ArgTypeList )}, as shown in Figure~\ref{fig:agent_pipeline}. 
LLMs cannot always generate accurate names of the code elements even if they have been provided in Step 2.
This issue is more common for open-source LLMs.
Therefore, the structured output of LLMs will be further refined using our postprocessing process, which matches the method names provided by LLMs to actual methods in the buggy program (see Section~\ref{section:postprocessing}). 
At last, the result is a refined and realistic list of probable buggy methods.

\begin{table*}[t]
    \centering
    \caption{Function calls designed for agents}
    \label{tab:function_calls}
    \begin{tabular}{l|l|l}
        \toprule
        \textbf{Name} & \textbf{Arguments} & \textbf{Description}\\
        \midrule
          get\_paths & None & Get the paths of the Java software system\\
        \midrule
          get\_classes\_of\_path & path\_name & Get the classes in the path of the Java software system\\
        \midrule
          get\_methods\_of\_class & class\_name & Get the methods belonging to the class of the Java software system\\
        \midrule
          get\_code\_snippet\_of\_method & method\_name & Get the code snippet of the Java method\\
        \midrule
          find\_class & class\_name & Find the class through fuzzy search\\
        \midrule
          find\_method & method\_name & Find the method through fuzzy search\\
        \midrule
          exit & None & Exit function calling\\
        \bottomrule
    \end{tabular}
\end{table*}

\subsubsection{Design of Function Calls}\label{section:function_call}

We design a set of function calls for the agents in \name to interact with, which assist LLMs in obtaining information related to bugs in the program.
These function calls need to know the structure and existing entities of the buggy program.
To meet this requirement, we traverse all the files in the program and extract the code snippets of all the methods using a parser.
For each method in the program, we record its fully qualified name, which contains its file path, class name, method name, and method signature.

We design seven function calls, which are listed and described in Table~\ref{tab:function_calls}.
The function calls beginning with \texttt{get} are intended to enable LLMs to navigate the file structure of the program or retrieve the code snippet of any specified method.
\new{Using \texttt{get\_paths}, \texttt{get\_classes\_of\_path}, and \texttt{get\_methods\_of\_class}, we prompt LLMs to explore bug-related methods step by step, from their belonging paths to their classes and then to themselves.
This design conserves substantial context tokens by omitting methods within paths and classes that LLMs are not focused on, and omitting the duplicated name prefixes of methods belonging to the same class or classes within the same path.
Taking the bug Time-25 for example, LLMs can utilize the function call \texttt{get\_classes\_of\_path(org.joda.time.tz)} to get class names such as \texttt{FixedDateTimeZone} without the path name prefix (i.e., org.joda.time.tz), enabling focused exploration of classes within the specified path and conserving context tokens.
In contrast, directly referencing all classes, such as \texttt{org.joda.time.tz.FixedDateTimeZone} and \texttt{org.joda.time.tz.CachedDateTimeZone}, consumes significant context and diverts attention away from related information.
}
All these four function calls beginning with \texttt{get} require fully qualified names (FQNs) of corresponding code elements as arguments.
However, LLMs often provide inaccurate or incomplete code entity names due to hallucinations~\cite{zhang2023siren, liu2024exploring} and limited capacity, which are more severe for open-source LLMs.
For instance, when calling function \texttt{get\_code\_snippet\_of\_method}, LLMs may only provide the name of the function without a signature or its belonging class, like \texttt{getOffsetFromLocal} shown in Table~\ref{tab:case_study}, which is incomplete to be precisely paired to any code snippet.
To enhance the utility and accuracy of the information gathered in each conversation, we match the incomplete or inaccurate names with actual code elements in the buggy programs using our postprocessing process (see Section~\ref{section:postprocessing}).
Specifically, when getting a function call from LLMs, we check if the code element specified by the given argument of the call exists in the buggy program.
If there is no code element named with the given argument, we call the postprocessing process shown in Algorithm~\ref{alg:Postprocess} with the argument and the fully qualified names of all counterparts in the buggy program.
For instance, we get the FQNs of all methods in the program to match with the given argument \texttt{getOffsetFromLocal}, which is provided by the LLM as a method's name for the function call \texttt{get\_code\_snippet\_of\_method}.
These function calls enable LLMs to autonomously get and understand adequate code information in the repository for fault localization.

Previous work~\cite{iFixR} finds that only a few files contain buggy methods, which account for a small fraction of the software system.
In addition, functions like \texttt{get\_classes\_of\_path} often return plenty of class names that are unrelated to the bug, thus consuming a significant portion of the context length and leaving limited tokens for LLMs to retrieve and inspect code snippets.
Considering that the text descriptions of both bug reports and trigger tests usually contain incomplete or fully qualified names of bug-related classes or methods, we design two functions that start with ``find``, which perform fuzzy searches to localize entities based on incomplete names and assist LLMs in obtaining the full qualified names of suspicious methods rapidly.
These function calls are implemented by passing the argument given by LLMs and the FQNs of counterparts to fuzzy search to our postprocessing process (see Section~\ref{section:postprocessing}).
With these functions for fuzzy search, LLMs can rapidly pinpoint entities mentioned in bug reports and trigger tests, which are likely related to bugs.

Additionally, as mentioned in Section~\ref{section:agent_pipeline}, there is a fixed function call \texttt{exit()}.
It allows LLMs to flexibly terminate Step 2 (i.e., Interaction with function calls) early via calling \texttt{exit()} to convey that they are confident in localizing bugs in Step 3 (see Section~\ref{section:agent_pipeline}).
This exit mechanism helps prevent LLMs from the interference of unrelated information extracted in redundant iterations when the number of loops is fixed.

\subsubsection{Postprocessing process}\label{section:postprocessing}

\begin{algorithm}[t]
  \caption{Postprocessing Process}
  \label{alg:Postprocess}
  \begin{algorithmic}[1]
    \Require
      The name of query $query$;
      The names of entities to search $entities$.
    \Ensure
      The matching names $names$.
    \State $names$ = []
    \State $query\_split$ = split($query$)
    \For{$entity$ in $entities$}
        \State $entity\_splits$ = split($entity$)
        \If{$query\_split$ all in $entity\_splits$}
            \State $names$.add($entity$)
        \EndIf
    \EndFor
    \If{length($names$) > 0}
        \Return $names$
    \EndIf
    \For{$entity$ in $entities$}
        \State $edit\_distance$ = Levenshtein.distance($query$, $entity$)
        \State $edit\_distances$.add($edit\_distance$)
        \If{$edit\_distance$ < 5}
            \State $names$.add($entity$)
        \EndIf
    \EndFor
    \If{length($names$) > 0}
        \Return $names$
    \EndIf
    \State \Return $names$[:5]
  \end{algorithmic}
\end{algorithm}

As mentioned before, we design a postprocessing process to match LLMs' inaccurate output with actual code elements in the buggy program, which helps provide more information in the limited context during the step of interaction with function calls and assists in pinpointing suspicious methods in the step of summarization.
The postprocessing process, as detailed in Algorithm~\ref{alg:Postprocess}, has two primary inputs: an inaccurate name provided by LLMs and the fully qualified names of all counterparts in the buggy program.
Specifically, the algorithm begins by splitting the query and the entity names using delimiters such as `.`, `/`, and `(`. 
The algorithm checks if the split components from the entity names contain all segments of the query. 
If a complete match is found, these names are returned as correct matches.
If no exact matches are found, the algorithm proceeds to measure the Levenshtein distance between the query and each entity name. 
The algorithm prioritizes entity names with a Levenshtein distance of less than five, based on empirical findings that this threshold works well for the types of names typically found in our scenario. 
If no names meet this threshold, the algorithm returns the five closest matches.

\subsubsection{\nameFL}
We designed an agent named \nameFL to narrow down the search space for FL.
\nameFL aims at global searching bug-related methods in the whole repository.
To achieve the goal, besides the basic function calls starting with ``get`` (e.g., \texttt{get\_path}) for obtaining code information, \nameFL also uses function calls starting with ``find`` (e.g., \texttt{find\_class}) for fuzzy searching in huge code space.
Following the pipeline in Section~\ref{section:agent_pipeline}, with the given input (i.e., bug reports and trigger tests) and descriptions of all the designing function calls, LLMs are prompted to analyze the bug and search globally for relevant information.
For example, for the bug Time-25 presented in Table~\ref{tab:input}, \nameFL first finds class \texttt{DateTimeZone} mentioned in the bug report and then gets its methods.
After extracting global information from the whole repository, \nameFL finally provides top-k suspicious methods for localization refinement. 

\subsubsection{\nameRefine}\label{sections:Agent4LR}
After obtaining the candidate list from space reduction, \name leverages another agent \nameRefine to perform a local exact search and refine the candidate list.
Different from \nameFL, we only provide \nameRefine with the \texttt{get\_code\_snippet\_of\_method} and \texttt{exit} function calls because we want this agent to save more attention and context length to focus on checking the code snippets of suggested methods in the given candidate list.
Through selecting suggested methods to scrutinize their code snippets, \nameRefine finally localizes buggy methods from the candidate list.
Since the fully qualified names of the suspicious methods are all given and open-source LLMs have difficulty retelling them, we call the function \texttt{get\_code\_snippet\_of\_method} with the index of the candidate method, helping \nameRefine better focus on double-checking.
In addition, we append the fully qualified name of the method that is double-checked via its index in the candidate list to the code snippet retrieved by the function call.
This can remind the LLMs of the connection between the code snippet and the fully qualified name of the method for the summarization step.

\section{Experimental Setting}
\label{section:setting}
\subsection{Research Questions}
\label{section:RQ}
We investigate the following research questions:
\begin{itemize}[leftmargin=*]
\item \textbf{RQ1: How does \name compare to other FL techniques?} 
We evaluate the effectiveness of \name by comparing it with leading FL techniques on the Defects4J benchmark.
\item \textbf{RQ2: How effective are the design choices in \name?} We compare the performance of \name with its variants on the Defects4J benchmark to illustrate the effectiveness of its components.
\item \textbf{RQ3: Can \name effectively localize bugs based on different open-source LLMs?} 
We investigate the generalizability of our \name by evaluating its performance on different open-source LLMs on the Defects4J benchmark.
\item \textbf{RQ4: Can \name effectively navigate and localize bugs in the wild?} 
We conduct experiments to evaluate the generalization capability of \name on the GHRB dataset where the projects are larger in size and the bugs are free from data leakage.
\end{itemize}

\subsection{Datasets}
Defects4J is a widely used benchmark for automatic fault localization, comprising a manually curated collection of real-world bugs from 17 Java projects~\cite{Defects4J, rafi2024exploring}. 
Defects4J (v2.0.0) includes a total of 835 bugs, among which all of the bugs are paired with developer-written trigger tests, and 814 bugs are paired with bug reports.

\subsection{Baselines}
To investigate the performance of \name, we select state-of-the-art FL techniques of different families as baselines:
\begin{itemize}[leftmargin=*]
    \item \textbf{Information Retrieval-based FL (IRFL):} We compare \name with  BoostNSift~\cite{BoostNSift} which is the state-of-the-art among method-level IRFL techniques that utilize merely bug reports.
    However, BoostNSift needs historical bug reports as input, which is not available in our input.
    To run it in our datasets, we remove the component of BoostNsift that requires historical bug reports, refer to the modified version as BoostN.
    \item \textbf{Spectrum-based FL (SBFL):} We consider two most commonly used formula-based SBFL techniques, i.e., Ochiai~\cite{Ochiai} and Dstar~\cite{Dstar}, which are also used by prior LLM-based FL studies~\cite{AutoFL,AgentFL}. 
    \item \textbf{Mutation-based FL (MBFL)}: We consider two representative MBFL techniques, MUSE \cite{Muse} and Metallaxis \cite{Metallaxis} following prior LLM-based FL studies~\cite{AutoFL}.
    \item \textbf{LLM-based FL}: We compare \name with existing method-level LLM-based FL techniques AutoFL~\cite{AutoFL} and AgentFL~\cite{AgentFL}.
    \item \textbf{HybridFL:} We compare \name with the state-of-the-art unsupervised technique SBIR~\cite{SBIR} that utilizes both bug reports and test cases to our best knowledge. 
    SBIR performs FL at the statement level. To compare \name with it, we transfer the ranked statement list produced by SBIR to the method level as follows. First, we replace each statement in the ranked list with the method it belongs to. If a statement is out of any method, i.e., a field declaration in a class, we remove it from the ranked list. Then, we scan the ranked list from the top and only collect a method when it is scanned for the first time to get the method-level rank.
    SBIR provides 10 results on 815 bugs with different random seeds that are used for the Monte Carlo algorithm~\cite{MonteCarlo} to combine its IRFL and SBFL ranks.
    We choose its FL results when the seed is set to 1 to compare \name with it.
\end{itemize}

\subsection{Evaluation Metric}
We use three evaluation metrics, i.e., Top-N, MAP, and MRR, which are widely used in the field of fault localization~\cite{BoostNSift,Muse,proFL,SBIR,AutoFL}.
The higher value of each metric represents better performance.

\textbf{Top-N:} Top-N computes the number of bugs with at least one buggy element appearing in the Top-N positions of the recommendation list. 
The Top-N metric has the additional benefit that it is a closer measure of what developers expect from FL \cite{expectations_FL}.
As suggested by prior work~\cite{apr_help}, usually, programmers only inspect a few buggy elements at the top of the given ranked list, e.g., 73.58\% developers only inspect Top-5 elements~\cite{expectations_FL}, we use Top-N (N=1, 3, 5).

\textbf{MAP:} Mean Average Precision (MAP) \cite{MAP} measures the average position of all the buggy methods 
localized by the bug localization method in the recommendation list. The definition is as follows:
\begin{align*}
    MAP &= \frac{1}{n} \sum_{j=1}^{n} AvgP_j\\
    AvgP_j &= \frac{1}{|K_j|} \sum_{k \in K_j} {Prec@k}\\
    Prec@k &= \frac{1}{k} \sum_{i=1}^{k} IsRelevant(i)
\end{align*}
Here, $AvgP_j$ is the average precision for the $j$-th bug,
and $|K_j|$ is the total number of buggy methods for the $j$-th bug. $Prec@k$ represents the precision of the top k methods in the recommendation list, and $IsRelevant(i)$ returns 1 if the $i$-th method in the recommendation list is responsible for the bug, and 0 otherwise. 
$K_j$ are ranks of buggy methods of the $j$-th bug.

\textbf{MRR:} Mean Reciprocal Rank (MRR) \cite{MRR} measures the position of the first buggy method localized by the bug localization method in the recommendation list. The definition is as follows:
\begin{align*}
    MRR = \frac{1}{n} \sum_{j=1}^{n} \frac{1}{rank_j}
\end{align*}
Here, $rank_j$ represents the ranking position of the first buggy method modified to fix the $j$-th bug in the recommendation list. 

\subsection{Implementation Details}\label{section:env}
\new{
We build \name based on the LLM Meta-Llama-3-8B-Instruct~\cite{llama3}, which is one of the state-of-the-art open-source LLMs and lightweight enough to conduct abundant experiments cheaply, rapidly, and greenly.
}
In order to make our experiments easy to reproduce, we set the temperature to 0.0 and top\_p to 1.0 in default.
In this work, we set the initial value of \textit{MAX} to 10 and the value of \textit{k} to 5 for our pipeline of agents (see Figure~\ref{fig:agent_pipeline}) following prior work~\cite{AutoFL}.
We set the value of \textit{m}, i.e., the size of the candidate list produced in the space reduction stage, to 20, due to the limited context length of LLMs.

In the space reduction stage, \name combines \nameFL with non-LLM-based FL techniques.
In our implementation of \name, we consider the non-LLM-based FL techniques as follows.
For IRFL approaches, we choose the SOTA method-level IRFL technique BoostN~\cite{BoostNSift}.
For SBFL approaches, we select Ochiai~\cite{Ochiai}, which is proven to be one of the most effective ranking strategies in object-oriented programs~\cite{FL_families} and is widely used by most FL tools that take test suites as input~\cite{PRFL,SBIR,Muse}.
Following prior work~\cite{SBIR}, we use GZoltar(v1.7.2)~\cite{GZoltar} to reproduce the FL results of Ochiai.
MBFL techniques need to modify all possible statements in the program and execute test cases multiple times, thus consuming hours to localize bugs~\cite{FL_families,AutoFL}.
Therefore, we do not use MBFL techniques~\cite{Muse,Metallaxis} for space reduction.
When bug reports and test cases are both available, we use the SOTA unsupervised HybridFL technique SBIR~\cite{SBIR} in the space reduction stage.
SBIR has been evaluated on the Defects4J dataset by Manish et al.~\cite{SBIR}, thus we directly utilize their evaluation results for experiments.
As mentioned, SBIR provides 10 results and we choose its FL results when the seed is set to 1 to ensure the reproducibility of our approach.
In this work, we also do not consider learning-based fault localization techniques since they need additional datasets for training, which are not available in unsupervised scenarios.

To combine \nameFL with non-LLM-based FL techniques and obtain m=20 suspicious methods, \name gets the top-5 ranks of SBIR, Ochiai, BoostN, and \nameFL successively in the localization refinement stage when bug reports and trigger tests are both available.
If only trigger tests are available, \name gets the top 15 suggested methods of Ochiai and the top 5 of \nameFL.
Similarly, \name gets the top 15 suggested methods of BoostN and the top 5 of \nameFL when only bug reports are available.

\section{Evaluation}
\label{section:evaluation}
In this section, we present our experimental results in detail with respect to the research questions introduced in Section~\ref{section:RQ}.

\subsection{RQ1. Overall Performance}
We evaluate the overall performance of our \name by comparing it with existing FL techniques on the Defects4J dataset and conduct qualitative analysis to illustrate why \name works.

\subsubsection{Quantitative Analysis:}
\label{section:RQ1.1}
Our \name is compared with existing FL techniques, both non-LLM-based FL and LLM-based FL techniques, to demonstrate its excellent capability of bug localization. 

\noindent\textbf{Evaluation on Defects4J (v2.0.0): }Table~\ref{tab:RQ1.1} presents a performance comparison between our \name and baselines (i.e., BoostN, Ochiai, and SBIR) in localizing buggy methods on the Defects4J (v2.0.0) dataset.
Evaluation results show that our \name successfully localizes 350 bugs within Top-1, 478 bugs within Top-3, and 529 bugs within Top-5, significantly surpassing all baseline techniques on the Defects4J (v2.0.0) benchmark.
This highlights the effectiveness of \name in fault localization. 
Also, the MRR and MAP values of \name are the best among all studied techniques, which demonstrate the high performance of \name in localizing multiple methods.

\begin{table}[t]
  \centering
  \caption{\name vs other FL techniques on Defects4J (v2.0.0)}
  \label{tab:RQ1.1}
  \resizebox{0.48\textwidth}{!}{
  \begin{tabular}{c|c|ccc|cc}
    \toprule
    Family & Technique  & Top-1 & Top-3 & Top-5 & MAP & MRR\\
    \midrule
    IRFL & BoostN & 149 & 241 & 280 & 0.206 & 0.235\\
    SBFL & Ochiai & 167 & 316 & 389 & 0.259 & 0.297\\
    HybridFL & SBIR & 222 & 377 & 433 & 0.309 & 0.362\\
    \midrule
    \multirow{2}{*} {LLM-based} & \name & \textbf{350} & \textbf{478} & \textbf{529} & \textbf{0.439} & \textbf{0.501}\\
    & \name+Repetition & \textbf{363} & \textbf{511} & \textbf{558} & \textbf{0.463} & \textbf{0.526}\\
    \bottomrule
  \end{tabular}
  }
\end{table}


\noindent\textbf{Evaluation on Defects4J (v1.0): }
When compared with LLM-based techniques, it is expensive to reproduce the results of AutoFL on all 835 bugs in Defects4J (v2.0.0), which requires API calling for GPT-3.5 and GPT-4.
Therefore, we compare \name to the performance of AutoFL reported in its paper~\cite{AutoFL}.
Note that AutoFL's evaluation is limited to 353 active bugs in five projects of Defects4J (v1.0) (i.e., Chart, Closure, Lang, Math, Time), which is a subset of Defects4J (v2.0.0).
Additionally, AutoFL does not use MAP and MRR as evaluation metrics.
To ensure a fair comparison, we also evaluate \name on the dataset used by AutoFL in the Top-N metrics.
Table~\ref{tab:RQ1.2} shows the comparative results of \name against AgentFL and other baselines (i.e., Muse, Metallaxis, Ochiai, DStar, and AutoFL) on all active bugs in Defects4J (v1.0) following AutoFL~\cite{AutoFL}.
Compared to MBFL and SBFL baselines, \name consistently localizes more bugs across all settings, including Top-1, Top-3, and Top-5. 
Built on the open-source LLM Llama3-8B, which not only ensures greater transparency of our approach but also enhances data security, \name outperforms LLM-based FL approaches AgentFL and AutoFL, both of which use GPT-3.5.
This superior performance highlights the effectiveness of \name in accurately identifying more bug locations with open-source LLMs.
AutoFL with GPT-4 achieves the best performance among all the methods.
However, the performance of \name in terms of Top-3 and Top5 is close to AutoFL-GPT-4.
It is also worth mentioning that GPT-4 is an expensive closed-source model, while the underlying model of \name is Llama3-8B, which is relatively smaller, cheaper to use, and can be self-hosted to address the concerns about data privacy.
\new{
Moreover, FlexFL can leverage AutoFL-GPT-4 as a standalone FL technique in its first stage. 
We evaluate a variant of FlexFL namely FlexFL+AutoFL-GPT-4 by combining the localization results of AutoFL-GPT-4, SBIR, Ochiai, and BoostN to produce a candidate list in the first step and utilize Agent4LR based on Llama3-8B-Instruct to refine the results in the second step.
As shown in Table 4, FlexFL+AutoFL-GPT-4 outperforms AutoFL-GPT-4 in both top-3 and top-5 scores and localizes 10 more bugs than AutoFL-GPT-4 at top-5, demonstrating that FlexFL as an effective framework can leverage and improve AutoFL. 
The performance decline in the top-1 metric is because Llama3-8B fails to precisely localize the complex bugs that GPT-4, with significantly stronger model capability, successfully resolves.
}
Therefore, we believe \name provides unique benefits compared to existing LLM-based FL techniques. 

\noindent\textbf{Fair comparison with GPT-3.5-based baselines: }
\new{
To comprehensively compare \name and GPT-3.5-based FL approaches, we also implement \name based on GPT-3.5. 
AgentFL and AutoFL used GPT-3.5-turbo-0613, which has been deprecated by OpenAI.
To ensure a fair comparison, we use the closest version to GPT-3.5-turbo-0613, i.e., GPT-3.5-turbo-1106, to reproduce the baselines and implement our FlexFL.
Unfortunately, AgentFL does not provide its implementation, so we can only replicate AutoFL.
As shown in Table~\ref{tab:RQ1.2}, FlexFL-GPT-3.5-1106 also outperforms AutoFL-GPT-3.5-1106 in terms of all the metrics and can localize 19 more bugs at top-5 with less than half the cost, demonstrating the effectiveness and economic efficiency of \name.
Note that, FlexFL based on Llama3-8B achieves better performance than FlexFL-GPT-3.5-1106 at top-1. 
After manual inspection of all the 39 bugs that FlexFL-Llama3-8B successfully localizes at top-1 while FlexFL-GPT-3.5 fails, we find that GPT-3.5 tends to be more confident and thus exits the phase of interaction with function calls earlier than Llama3-8B, resulting in insufficient information for precise localization. 
For example, in the second stage, FlexFL-GPT-3.5 calls an average of only 2.82 functions, while FlexFL-Llama3-8B calls 4.56 functions.
Moreover, we also adapt AutoFL to use Llama3-8B by modifying its original function call API provided by OpenAI GPT to match the one used in FlexFL. 
Evaluation results in Table~\ref{tab:RQ1.2} demonstrate that AutoFL-Llama3 performs significantly worse than AutoFL-GPT-3.5, localizing 41 fewer bugs at top-1 for example. 
These results underscore the adaptability of FlexFL with lightweight open-source LLMs. 
}

\noindent\textbf{Investigation on repetition strategy: }
\new{
AutoFL~\cite{AutoFL} repeats its pipeline multiple runs and aggregates the results, which can also be applied to \name.
To this end, we set the temperature to 0.6 and top\_p to 0.9 to make Llama3-8B-Instruct output stochastically, following the default setting in its model card~\cite{Scripts-Llama3}.
AutoFL gives a score of $1/n$ to each of the identified methods if a prediction contains $n$ methods. 
These individual scores are then averaged over all $R$ predictions. 
Different from AutoFL, which does not rank suspicious methods in each run, \name, i.e., \nameFL and \nameRefine, localize and rank the top k most suspicious methods.
Therefore, we added a factor that reflects the ranks of the identified methods to the formula of AutoFL. Formally, the score of a method $m$ is defined as: 
\begin{align*}
    score(m) = \frac{1}{R} \sum_{i=1}^{R}( \frac{1}{|r_i|} \cdot [m \in r_i] \cdot \frac{1}{rank_i})
\end{align*}
where $r_i$ is the set of predicted methods from the $i$-th run, $[.]$ is the Iverson bracket which returns 1 when the predicate inside is true and 0 otherwise, and $rank_i$ is the rank of method $m$ in $r_i$. 
We first repeat to run \nameFL $R$=5 times following AutoFL-GPT-3.5 and aggregate the results using our formula. 
Then we aggregate the results of all the runs and produce a candidate list for the second stage.
Finally, we also repeat to run \nameRefine $R$=5 times with the new candidate list and aggregate the results. 
We refer to this variant of FlexFL as FlexFL+Repetition. 
As shown in Table~\ref{tab:RQ1.1} and Table~\ref{tab:RQ1.2}, FlexFL+Repetition can further improve \name. Specifically, it localizes 29 and 9 more bugs at top-5 than the original \name on Defects4J (v2.0.0) and Defects4J (v1.0), respectively, demonstrating that FlexFL can also benefit from the repetition strategy.
}

\begin{table}[t]
  \centering
  \caption{\name vs vs other FL techniques on Defects4J (v1.0)}
  \label{tab:RQ1.2}
  \begin{threeparttable}
  \resizebox{0.5\textwidth}{!}{
  \begin{tabular}{c|c|ccc}
    \toprule
    Family & Technique  & Top-1 & Top-3 & Top-5\\
    \midrule
    \multirow{2}{*} {MBFL} & MUSE & 73 & 139 & 161 \\
    & Metallaxis & 106 & 162 & 191 \\
    \midrule
    \multirow{2}{*} {SBFL} & Ochiai & 122 & 192 & 218 \\
    & DStar & 125 & 195 & 216 \\
    \midrule
    \multirow{9}{*} {LLM-based} & AgentFL & 144 & 169 & 173 \\
    & AutoFL-Llama3-8B & 105 & 157 & 172 \\
    & AutoFL-GPT-3.5\tnote{\#} & 146 & 180 & 194\\
    & AutoFL-GPT-3.5-1106 (\$22.26)\tnote{*} & 151 & 209 & 221\\
    & \textbf{\name-GPT-3.5-1106 (\$9.71)}\tnote{*} & 154 & 220 & 240 \\
    & \textbf{\name} & 164 & 214 & 236 \\
    & \textbf{\name+ Repetition} & \textbf{164} & \textbf{220} & \textbf{245} \\
    \cline{2-5}
    \noalign{\smallskip}
    & AutoFL-GPT-4 & \textbf{187} & 236 & 251 \\
    & \textbf{\name+ AutoFL-GPT-4} & 176 & \textbf{239} & \textbf{261} \\
    \bottomrule
  \end{tabular}
  }
  \begin{tablenotes}
        \footnotesize
        \item[\#] used GPT-3.5-turbo-0613(deprecated) by AutoFL in its paper.
        \item[*] use GPT-3.5-turbo-1106 for reproduction.
      \end{tablenotes}
  \end{threeparttable}
\end{table}

\subsubsection{Overlap of Different Methods:}
\label{section:overlap}
\begin{figure}[t]
\centering
\subfigure[]{
\includegraphics[width=0.22\textwidth]{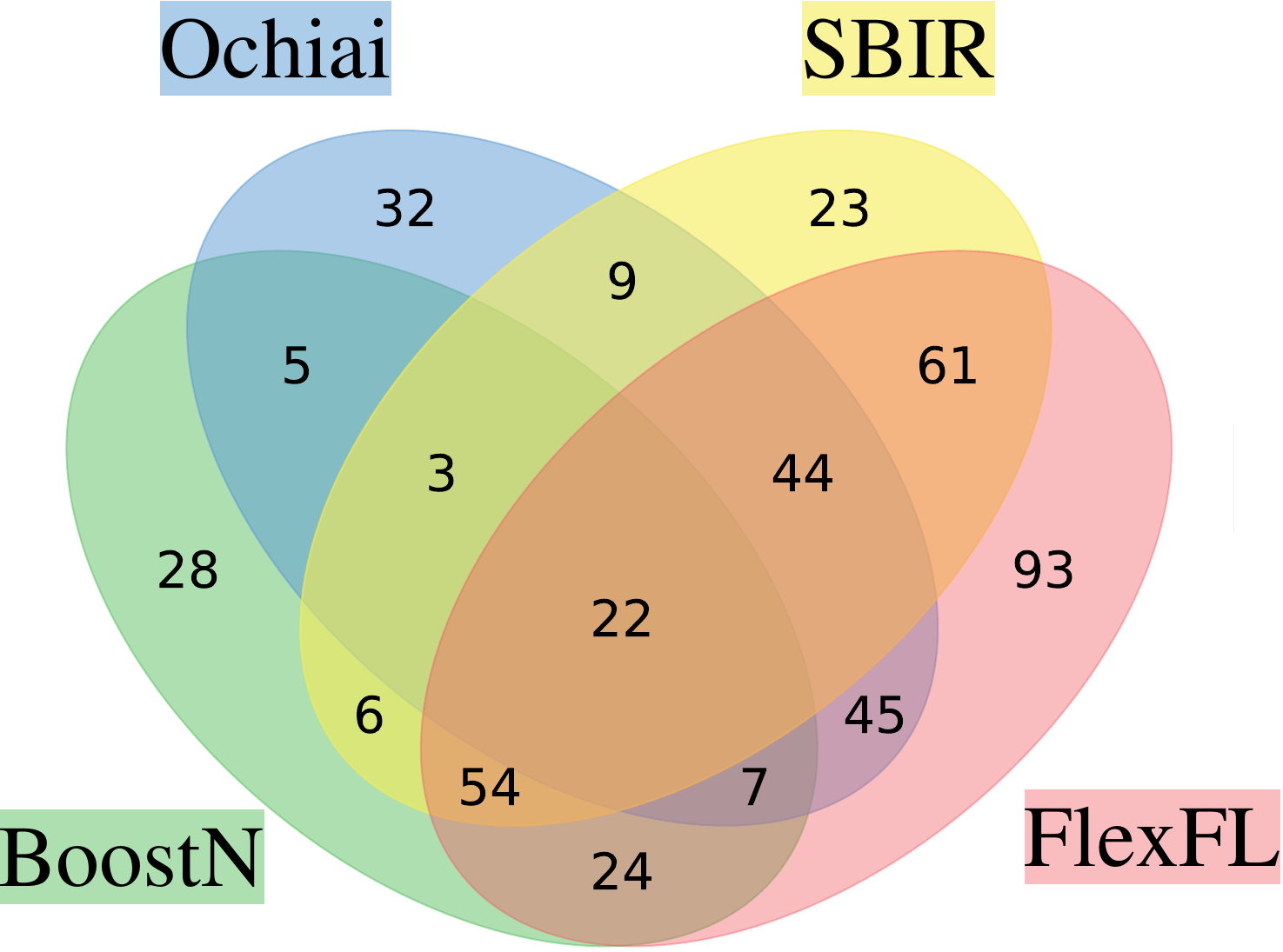}
\label{fig:venn_a}
}
\subfigure[]{
\includegraphics[width=0.20\textwidth]{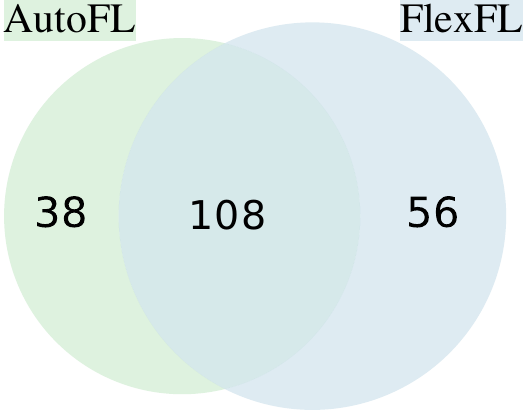}
\label{fig:venn_b}
}
\caption{Overlap Analysis of \name and (a) non-LLM-based FL techniques (b) LLM-based FL approaches}
\label{fig:venn}
\end{figure}

We analyze the overlap of bugs localized in Top-1 between \name and non-LLM-based FL techniques on Defects4J (v2.0.0).
As shown in Figure~\ref{fig:venn_a}, \name successfully localizes 93 bugs that other non-LLM-based methods miss at Top-1. This demonstrates that \name can complement and enhance existing non-LLM-based FL approaches, rather than just combining their results.
\new{We extend the analysis to compare \name with LLM-based baselines.
Since AgentFL does not provide its results or implementation, we can only analyze the overlapping bugs localized in Top-1 on Defects4J (v1.0) between \name and AutoFL-GPT-3.5.
As shown in Figure~\ref{fig:venn_b}, \name can localize 56 bugs that AutoFL-GPT-3.5 misses at Top-1 while AutoFL-GPT-3.5 only localizes 38 unique bugs, demonstrating the effectiveness of \name.
After inspecting these cases, we find that FlexFL has two key advantages over AutoFL: 
(1) FlexFL can leverage bug reports while AutoFL cannot. 
For example, the bug report of Closure-113 clearly identifies the buggy method in its description: “ProcessClosurePrimitives pass has a bug in processRequireCall method”. 
Thus both Agent4SR and Agent4LR in FlexFL localize this bug without effort with the help of the bug report.
(2) FlexFL integrates results from non-LLM-based FL approaches to reduce search space while AutoFL relies solely on LLMs to localize bugs. 
For example, the bug Math-38 is localized at top-1 by SBIR, which assists FlexFL in performing a successful localization. 
In contrast, GPT-3.5 in AutoFL struggles to identify the buggy method among numerous classes and methods covered by the trigger test.
}

\new{
However, some bugs are missed by FlexFL but localized by other FL techniques, as shown in Figure~\ref{fig:venn}. 
Theoretically, FlexFL has the potential to successfully localize these bugs since one of BoostN, Ochiai, and SBIR has localized the buggy methods at top-1 and included them in the candidate list. 
However, constrained by the limitations of its base model, FlexFL struggles to effectively reason about and comprehend the bug-related information and the code snippets of the methods in the candidate list, failing to address these bugs. 
For example, the bug Math-4 is localized at top-1 by SBIR, Ochiai, and BoostN. 
However, Llama3-8B in FlexFL insists that “NullPointerException is thrown in the Line.getAbscissa method” and thus localizes it at top-1. 
Nonetheless, FlexFL also offers a reasoning chain for examining the methods in the candidate list in the second stage, which can, to some extent, assist human debuggers in further reviewing the results of FlexFL and the technologies used in its first stage. 
On the other hand, AutoFL provides LLMs with function calls that get classes and methods covered by trigger tests, which helps LLMs reduce search space and thus localize additional bugs. 
Take the bug Lang-38 for example. 
AutoFL first narrows the search space to the methods and classes that are covered by the trigger test and then successfully identifies the buggy method \texttt{org.apache.commons.lang3.time.FastDateFormat.
format(Calendar,StringBuffer)} by checking the suspicious methods one by one. 
However, both Agent4SR and Agent4LR are misled by the incorrect class name \texttt{DateFormatUtils} provided in the bug report, and non-LLM-based FL approaches we use in the first stage also fail to locate the buggy method at the top-5. 
Nonetheless, we find that FlexFL can successfully localize this bug by integrating the result of AutoFL-GPT-3.5 in the first stage, indicating that this limitation can be resolved by incorporating a broader range of powerful FL approaches during the space reduction stage of FlexFL. 
}


\subsubsection{Case study}
\begin{table*}[hbtp]
    \centering
    \caption{Case Study of Example Bug: Time-25 in Defects4J}
    \label{tab:case_study}
    \begin{tabular}{l|l}
        \toprule
        \textbf{Buggy method} & org.joda.time.DateTimeZone.getOffsetFromLocal(long)\\
        \midrule
        \multirow{9}{*}{\textbf{\nameFL}} & \textbf{Reason : }
        Based on the bug report and the trigger test, I can identify the following:\\
        & 1. The bug is related to the DateTimeZone.getOffsetFromLocal method, \\
        & which is responsible for calculating the offset from local time to UTC.\\
        & 2. The bug is triggered during a DST (Daylight Saving Time) transition, where the offset changes from -7 to -8.\\
        & 3. The trigger test is related to the Moscow timezone, which also has a DST transition.\\
        & \textbf{Function Calls : } 1. find\_class("DateTimeZone") 2. get\_methods\_of\_class("org.joda.time.DateTimeZone")\\
        & 3. get\_code\_snippet\_of\_method("getOffsetFromLocal")……\\
        & \textbf{Top\_1 :} org.joda.time.tz.FixedDateTimeZone.getOffsetFromLocal(long)\\
        & \textbf{Top\_3 :} org.joda.time.tz.DefaultNameProvider.getOffsetFromLocal(long)\\
        \midrule
        \multirow{7}{*}{\textbf{\nameRefine}} & \textbf{Reason : }
        Based on the bug report and the trigger test, I can identify the following key points:\\
        & 1. The bug is related to the DateTimeZone.getOffsetFromLocal method, specifically during DST transitions.\\
        & 2. The test case provided shows that the method is returning the wrong offset for a specific date and time.\\
        & 3. The trigger test is failing, indicating that the method is not handling the DST transition correctly.\\
        & \textbf{Function Calls : } 1.get\_code\_snippet\_of\_method(1) (i.e., org.joda.time.DateTime.DateTime(long,DateTimeZone))\\
        & 2.get\_code\_snippet\_of\_method(18) (i.e., org.joda.time.DateTimeZone.getOffsetFromLocal(long))……\\
        & \textbf{Top\_1 :} org.joda.time.DateTimeZone.getOffsetFromLocal(long)\\
        \bottomrule
    \end{tabular}
\end{table*}
To further investigate the performance of our \name and illustrate why it can work, 
we conduct a case study using the bug detailed in Table~\ref{tab:input}.
The buggy method is ranked 9th, 53th, and 9th by BoostN, Ochiai, and SBIR, respectively.
However, our \name can rank it within the top 1. 
The following details illustrate how \name effectively localizes the buggy method.

In the first stage of \name, based on the bug report, \nameFL reasons that \textit{this bug is related to DateTimeZone.getOffsetFromLocal method, which is responsible for calculating the offset from local time to UTC.}, as shown in Table~\ref{tab:case_study}.
According to this clue, \nameFL finds the class \texttt{DateTimeZone} and gets the code snippet of method \texttt{getOffsetFromLocal} with the assistance of function calls.
Then our function call returns the FQN of the class \texttt{DateTimeZone}, i.e., \texttt{org.joda.time.DateTimeZone}, and the methods named \texttt{getOffsetFromLocal}, which are in the \texttt{org.joda.time.DateTimeZone} class and the \texttt{org.joda.time.tz.FixedDateTimeZone} class.
However, due to the limited performance of the used open-source LLM, \nameFL is confused by the word \textit{Fixed} in the second method name and thus localizes this wrong method at the top 1.
In addition, \nameFL provides a method that does not exist in the buggy program in the third place, probably due to the hallucination~\cite{zhang2023siren} of the open-source LLM.
In the summarization stage of \nameFL, our postprocessing process is used to further match each predicted name with the FQNs of all the methods in the buggy program.
The FQN of the buggy method, i.e., \texttt{org.joda.time.DateTimeZone.getOffsetFromLocal\newline(long)}, has the minimum editing distance from the third predicted name.
Thus, \nameFL localizes the buggy method in the third place.

In the second stage of \name, \nameRefine automatically checks the suspicious methods localized by SBIR, Ochiai, BoostN, and \nameFL, which are arranged in order, to localize the buggy method.
\new{
In this case study, thanks to the language understanding and reasoning capabilities of LLMs, \nameRefine successfully identifies that "the bug is related to the DateTimeZone.getOffsetFromLocal method" from the title of the bug report shown in Table~\ref{tab:input}. 
Then, based on the reasoning results and the candidate list, Agent4LR generates the function call get\_code\_snippet\_of\_method(18) to check the code snippet of the truly buggy method \texttt{org.joda.time.DateTimeZone.getOffsetFromLocal\newline(long)} in detail.
Finally, after comprehending the retrieved code and reasoning about the bug, \nameRefine succeeds in locating this buggy method at top-1, demonstrating the benefits of \nameRefine.
}

In conclusion, different FL techniques used in the space reduction stage, including LLM-based \nameFL and non-LLM-based approaches, complement each other and suggest methods comprehensively via extracting information from bug reports and trigger tests in various ways.
In the localization refinement stage, \nameRefine can refine the result of the first stage and perform more accurate localization.
In addition, the function calls and the postprocessing process contribute to assisting agents in efficiently obtaining information relevant to bugs from the huge code base of the whole repository.

\begin{tcolorbox}[breakable,width=\linewidth-2pt,boxrule=0pt,top=2pt, bottom=2pt, left=2pt,right=2pt, colback=gray!20,colframe=gray!20]
\textbf{Answer to RQ1:}
Our \name outperforms existing fault localization techniques on Defects4J v1.0 and v2.0.0. Moreover, it can localize 93 unique bugs that cannot be localized by the non-LLM-based FL techniques used in the space reduction stage. 
\end{tcolorbox}

\subsection{RQ2. Ablation Study}
We conduct comprehensive ablation studies to investigate the impacts of different inputs and design choices on the effectiveness of \name.

\subsubsection{Ablation of Inputs}
\label{section:input_ablation}
\begin{table}[t]
  \centering
  \caption{Ablation study on input}
  \label{tab:RQ2.1}
   \resizebox{0.48\textwidth}{!}{
    \begin{tabular}{c|ccc|cc}
    \toprule
    Variant & Top-1 & Top-3 & Top-5 & MAP & MRR\\
    \midrule
    w/o trigger test  & 257 & 345 & 387 & 0.323 & 0.365\\
    w/o bug report & 266 & 395 & 442 & 0.339 & 0.399\\
    w/o buggy program & 45 & 64 & 75 & 0.062 & 0.066\\
    \midrule
    \name & \textbf{350} & \textbf{478} & \textbf{529} & \textbf{0.439} & \textbf{0.501}\\
    \bottomrule
  \end{tabular}
  }
\end{table}
We first compare the roles of different inputs in \name.
From Table~\ref{tab:RQ2.1}, we can observe that \name, which takes both bug reports and test cases as input, significantly outperforms \name w/o trigger test and \name w/o bug reports.
By flexibly leveraging available bug-related information, \name can localize at least 31.6\% more bugs at top-1, which indicates the effectiveness of flexibility in handling inputs that the existing LLM-based FL techniques lack.

We also evaluate the performance of the basic variant of \name that is not provided with function calls to access any information in the buggy program and directly localizes bugs with bug reports and test cases.
Table~\ref{tab:RQ2.1} shows that \name without buggy program can only localize 75 bugs at top-5 even with suspicious methods localized by the state-of-the-art non-LLM-based FL techniques, demonstrating that the tool use greatly improves the performance of \name and the impact of model memorization is limited.

\subsubsection{Ablation of Design Choices}
\label{section:ablation_study}
We investigate the contributions of the two-stage design and different design choices for assisting agents to better interact with function calls.

\noindent\textbf{Two-stage Design: } 
\name is composed of two stages, i.e., space reduction and localization refinement, and we design two agents for each stage, i.e., \nameFL and \nameRefine.
The localization refinement stage requires the candidate list produced by the space reduction stage.
Therefore, we cannot entirely remove the space reduction stage and thus focus on the effectiveness of combining different FL techniques in this stage.
We construct multiple variants of \name, each of which only uses one FL technique to produce top-20 suspicious methods in the first stage, and compare them with \name.
To investigate the effectiveness of the localization refinement stage, we compare these variants with the FL technique they use.
In addition, to investigate the contribution of \nameFL, we construct a variant by removing \nameFL from the first stage, i.e., \name without \nameFL, and compare it with \name.
\new{
Moreover, we also compare the performance of \nameFL and \name to better understand the contribution of our two-stage framework.
}

\begin{table}[t]
  \centering
  \caption{Ablation study on two-stage}
  \label{tab:RQ2.2}
   \resizebox{0.48\textwidth}{!}{
    \begin{tabular}{c|ccc|cc}
    \toprule
    Variant & Top-1 & Top-3 & Top-5 & MAP & MRR\\
    \midrule
    BoostN & 149 & 241 & 280 & 0.206 & 0.235\\
    BoostN + \nameRefine & 255 & 336 & 364 & 0.317 & 0.356\\
    \midrule
    Ochiai & 167 & 316 & 389 & 0.259 & 0.297\\
    Ochiai + \nameRefine & 303 & 421 & 475 & 0.387 & 0.442\\
    \midrule
    SBIR & 222 & 377 & 433 & 0.309 & 0.362\\
    SBIR + \nameRefine & 319 & 429 & 473 & 0.400 & 0.453\\
    \midrule
    \nameFL & 232 & 318 & 348 & 0.293 & 0.333\\
    \name w/o \nameFL & 338 & 467 & 510 & 0.422 & 0.485\\
    \name & \textbf{350} & \textbf{478} & \textbf{529} & \textbf{0.439} & \textbf{0.501}\\
    \bottomrule
  \end{tabular}
  }
\end{table}

Table~\ref{tab:RQ2.2} shows the evaluation results.
First, \name can achieve the best performance compared to all these variants, suggesting the effectiveness of the space reduction stage.
Second, the variants that leverage \nameRefine, e.g., SBIR + \nameRefine, outperform their used non-LLM-based FL techniques, e.g., SBIR, indicating that \nameRefine, as well as the localization refinement stage, can improve non-LLM-based FL techniques of various families. 
For instance, \nameRefine refines the results of Ochiai and increases its top-1 score by 81.4\%.
Third, \name outperforms \name without \nameFL, demonstrating that the synergy between \nameFL and non-LLM-based FL techniques can complement each other’s advantages and enhance bug localization.
\new{
In addition, \nameFL outperforms all the non-LLM-based baselines, i.e., BoostN, Ochiai, and SBIR, in the top-1 score, demonstrating its effectiveness as a standalone FL technique.
Specifically, \nameFL with text descriptions of bug-related information can localize 65 more bugs than Ochiai at top-1, which can leverage the dynamical spectrum collected by test suites.
Moreover, by integrating Agent4SR with non-LLM-based FL approaches in the first stage and refining their hybrid results in the second stage, 
FlexFL significantly improves Agent4SR across all settings and localizes 50.9\% (118) more bugs at top-1. 
These results demonstrate the superior effectiveness of our two-stage framework in enhancing FL performance.
}
To conclude, our two-stage design combines and refines the results of LLM-based and non-LLM-based FL approaches, which helps \name improve existing FL techniques.

\noindent\textbf{Designs for better interaction with function calls: } 
To release the power of open-source LLMs, \name designs three components to assist them in better interaction with function calls to extract information from the buggy program effectively.
First, \name prompts open-source LLMs to reason and plan how to use function calls before interacting with them.
Second, \name designs a postprocessing process to match inaccurate names provided by open-source LLMs to actual code elements in the repository.
Third, \nameRefine adapts the parameter of \texttt{get\_code\_snippet\_of\_method} to the method's number in the candidate list and append the FQN of the method checked to its code snippet.
To prove the usefulness of these three designs, we evaluate one variant of \name that interacts with function calls directly without reasoning, one variant that does not use the postprocessing process, and another that uses the same parameter as \nameFL to use \texttt{get\_code\_snippet\_of\_method} in the second stage, namely w/o focus.

\begin{table}[t]
    \centering
  \caption{Ablation study on designs for better tool use}
  \label{tab:RQ2.3}
   \resizebox{0.48\textwidth}{!}{
    \begin{tabular}{c|ccc|cc}
    \toprule
    Variant & Top-1 & Top-3 & Top-5 & MAP & MRR\\
    \midrule
    w/o reasoning & 330 & 466 & 512 & 0.418 & 0.480\\
    w/o postprocessing & 330 & 462 & 512 & 0.417 & 0.478\\
    w/o focus & 332 & 440 & 479 & 0.401 & 0.464\\
    \midrule
    \name & \textbf{350} & \textbf{478} & \textbf{529} & \textbf{0.439} & \textbf{0.501}\\
    \bottomrule
  \end{tabular}
  }
\end{table}

Results are shown in table~\ref{tab:RQ2.3}.
Compared to these variants, \name can localize at most 50 more bugs at top-5 and improve up to 8\% on both MAP and MRR metrics,
indicating the effectiveness of our designs for assisting open-source LLMs in better interaction with function calls.

\begin{tcolorbox}[breakable,width=\linewidth-2pt,boxrule=0pt,top=2pt, bottom=2pt, left=2pt,right=2pt, colback=gray!20,colframe=gray!20]
\textbf{Answer to RQ2:}
All design choices of \name make contributions to the whole framework's performance, the most significant of which is the flexibility of making use of different types of input, which improves \name at least 31.6\% on the top-1 metric.
\end{tcolorbox}

\subsection{RQ3. Generalizability across Open-Source LLMs}
To investigate the generalizability of \name across different open-source LLMs, we also implement \name based on two other lightweight open-source LLMs, i.e., Qwen2-7B-Instruct~\cite{Qwen2} and Mistral-Nemo-12B-Instruct~\cite{Mistral-Nemo}, respectively, and evaluate the two variants on Defects4J (v2.0.0).
Qwen2-7B and Mistral-Nemo-12B are famous lightweight open-source models and achieve state-of-the-art performances across various tasks~\cite{Open-source-LLMs-Leaderboard}.
Table~\ref{tab:RQ3} presents the evaluation results.
The variants of \name based on different LLMs achieve slightly different performances, indicating that the base model has an impact on the effectiveness of \name.
Among these variants, \name-Llama3-8B achieves the best performance in all metrics.
Additionally, the performance of \name-Qwen2-7B and \name-Mistral-Nemo-12B is comparable to \name-Llama3-8B, which indicates that \name can generally work with different lightweight open-source LLMs.

\begin{table}[t]
  \centering
  \caption{Comparison of \name based on different open-source LLMs on Defects4J (v2.0.0)}
  \label{tab:RQ3}
  \begin{tabular}{c|ccc|cc}
    \toprule
    Base Model  & Top-1 & Top-3 & Top-5 & MAP & MRR\\
    \midrule
    Mistral-Nemo-12B & 322 & 464 & 507 & 0.411 & 0.473\\
    Qwen2-7B  & 329 & 456 & 500 & 0.408 & 0.474\\
    Llama3-8B & \textbf{350} & \textbf{478} & \textbf{529} & \textbf{0.439} & \textbf{0.501}\\
    \bottomrule
  \end{tabular}
\end{table}

\begin{tcolorbox}[breakable,width=\linewidth-2pt,boxrule=0pt,top=2pt, bottom=2pt, left=2pt,right=2pt, colback=gray!20,colframe=gray!20]
\textbf{Answer to RQ3:}
\name demonstrates good generalizability in working with different lightweight open-source LLMs.
\end{tcolorbox}

\label{section:RQ3}

\subsection{RQ4. Generalization Capability on GHRB}~\label{sec:rq4}
\new{ 
Considering the LLM used in \name (see~\cref{section:env}) was trained with data collected until March 2023~\cite{cutoff_date} and the bugs in Defects4J are fixed before this time, performing evaluations on the Defects4J dataset may lead to concerns about data leakage. 
To mitigate such concerns, we also conduct experiments on the GHRB dataset~\cite{GHRB}, which was recently collected from 17 high-quality GitHub repositories that use JUnit. 
We evaluate \name and compare it to the state-of-the-art baselines of different families on a subset GHRB where the bugs are fixed after the training data cutoff point (i.e., March 2023) of Llama3 and that of GPT-3.5 (i.e., September 2021).
This subset contains 38 bugs, among which only 28 bugs are artificially reproducible.
We cannot reproduce the other 10 bugs because of the immaturity of GZoltar, which is required by Ochiai, SBIR, and AutoFL to collect dynamic coverage of programs. 
For example, even the latest version (v1.7.3) of GZoltar has not supported JUnit 5 which is used by some projects in the GHRB subset.
Experiments on this GHRB subset help investigate the generalization of \name to the latest large software systems.
Table~\ref{tab:datasets} shows the average number of source code files, faulty methods, and code lines within the buggy programs in the GHRB subset and Defects4J (v2.0.0).
The buggy programs in the GHRB subset are notably larger in size than those found in Defects4J, which to some extent help investigate the scalability of our approach in large-scale real-world projects.
}

\begin{table}[t]
  \centering
  \caption{Complexity of GHRB and Defects4J (v2.0.0)}
  \label{tab:datasets}
  \begin{threeparttable}
  \begin{tabular}{cccc}
    \toprule
    Dataset  & Files & Faulty Methods & LoC\tnote{*}\\
    \midrule
    Defects4J (v2.0.0) & 267 & 1.73 & 76.3k\\
    GHRB  & 1115 & 2.75 & 177.0k\\
    \bottomrule
  \end{tabular}
  \begin{tablenotes}
    \footnotesize
    \item[*] LoC is the average number of code lines for each buggy program.
  \end{tablenotes}
  \end{threeparttable}
\end{table}

\new{
Table~\ref{tab:RQ4} shows the evaluation results of \name, BoostN, Ochiai, SBIR, and AutoFL on the 28 reproducible bugs. 
The results show that both \name and \nameFL outperform all the baselines in terms of Top-1, Top-3, and Top-5, demonstrating their generalization capabilities for large projects in real cases.
Moreover, \name achieves the best performance, confirming the effectiveness of the synergy between \nameFL and non-LLM-based FL techniques.
Moreover, \name localizes 19 out of 28 bugs (i.e., 67.9\%) in the GHRB subset at top-1. The proportion is even higher than that on Defects4J (v2.0.0) (i.e., 41.9\%). 
These results reveal the generalization capability of \name on the GHRB subset which is free from data leakage concerns.
}

\begin{table}[t]
  \centering
  \caption{Comparison of \name with Baselines on GHRB}
  \label{tab:RQ4}
  \begin{tabular}{c|ccc|cc}
    \toprule
    Technique  & Top-1 & Top-3 & Top-5 & MAP & MRR\\
    \midrule
    BoostN  & 3 & 6 & 7 & 0.102 & 0.158\\
    Ochiai & 4 & 12 & 15 & 0.273 & 0.297\\
    SBIR & 12 & 15 & 18 & 0.434 & 0.505\\
    AutoFL-Llama3-8B & 12 & 15 & 15 & / & / \\
    AutoFL-GPT-3.5 & 13 & 16 & 16 & / & / \\
    \nameFL-Llama3-8B  & 15 & 18 & 20 & 0.557 & 0.601\\
    \name-GPT-3.5 & 16  & 20 & 21 & 0.570 & 0.644\\
    \name-Llama3-8B & \textbf{19} & \textbf{22} & \textbf{22} & \textbf{0.648} & \textbf{0.732}\\
    \bottomrule
  \end{tabular}
\end{table}

\begin{tcolorbox}[breakable,width=\linewidth-2pt,boxrule=0pt,top=2pt, bottom=2pt, left=2pt,right=2pt, colback=gray!20,colframe=gray!20]
\textbf{Answer to RQ4:}
FlexFL can generalize to the GHRB dataset where the projects are larger in size and the bugs are free from data leakage.
\end{tcolorbox}

\section{Discussion}
\label{section:discussion}
This section discusses the failure cases of \nameFL, the comparison between \name and learning-based FL techniques, and the threats to validity of this work.

\subsection{Qualitative Analysis of \nameFL's Failures}
\new{
As we have mentioned in Section~\ref{section:ablation_study}, \nameFL can be used as a standalone LLM-based FL approach and achieves good performance on Defects4J (v2.0.0). 
However, in many cases, Agent4SR could not make a successful FL. To understand the limitations of Agent4SR and provide insights for future work, we randomly sample and manually analyze 50 cases where Agent4SR fails to localize the bugs at top-5.}

\new{
The most common cause of failures, observed in 38 cases, is the need for more efficient approaches to assist LLMs in repository-level code exploration: while bug reports and trigger tests typically help clarify the bug behaviors, Agent4SR is limited to starting from classes or methods directly mentioned in the bug-related information or those with semantically related names, rather than those that are functionally relevant. 
Table~\ref{tab:failure_analysis} shows the bug report and trigger test of the bug Cli-2 in Defects4J, which have been lightly edited for clarity and space-saving.
Based on the description of the bug report, Agent4SR first identifies the bug behavior “The problem occurs when a parameter value contains a hyphen as the first character, which is misinterpreted as a parameter”. 
However, due to its unfamiliarity with the buggy program, Agent4SR is unable to directly pinpoint the methods that cause the bug behavior and thus can only start from the class \texttt{PosixParser}, which is explicitly mentioned in the trigger test. 
Then, Agent4SR chooses to check the code snippets of three methods within this class, e.g., \texttt{processOptionToken}, which appear suspicious but are innocent, failing to localize this bug. 
This case highlights a need to effectively incorporate project-specific information for LLMs. 
Incorporating search engines which can retrieve methods that implement related functionalities mentioned in bug-related information, or function calls that provide call relationships between methods can help alleviate this limitation. 
Other failure reasons included (1) low-quality bug-related information (3 cases), making it difficult even for a human debugger to localize the bug; (2) fixes applied outside methods, e.g., to a field of a class, while Agent4SR aims to find buggy methods (5 cases); (3) LLM errors in generating function calls, resulting in insufficient information for FL (2 cases); and (4) inadequate relevant information gathered from too few function calls due to context length limitations (2 cases).
}

\begin{table}[t]
    \centering
    \caption{Failure Analysis of Example Bug: Cli-2 in Defects4J}
    \label{tab:failure_analysis}
    \begin{tabular}{p{8.5cm}}
        \toprule
        \textbf{Bug report:} Title:[cli] Parameter value "-something" misinterpreted as a parameter\\
        Description:If a parameter value is passed that contains a hyphen as the (delimited) first character, CLI parses this a parameter......\\
        \midrule
        \textbf{Trigger test:} public void test() throws Exception\{\\
        \textit{1} Options options = buildCommandLineOptions();\\
        \textit{2} CommandLineParser parser = new PosixParser();\\
        \textit{3} String[] args = new String[] {"-t", "-something" };\\
        \textit{4} CommandLine commandLine=parser.parse( options, args );\\
        \bottomrule
    \end{tabular}
\end{table}

\subsection{Comparison with Learning-Based FL techniques}
\begin{figure}[t]
\centering
\subfigure[]{
\includegraphics[width=0.26\textwidth]{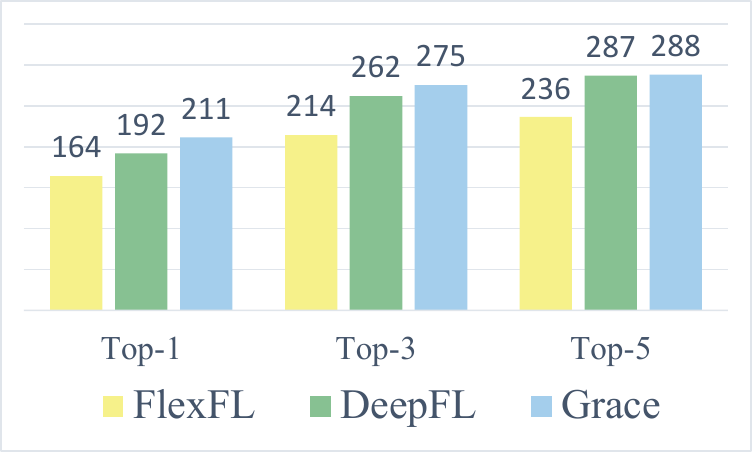}
\label{fig:table_LBFL}
}
\subfigure[]{
\includegraphics[width=0.16\textwidth]{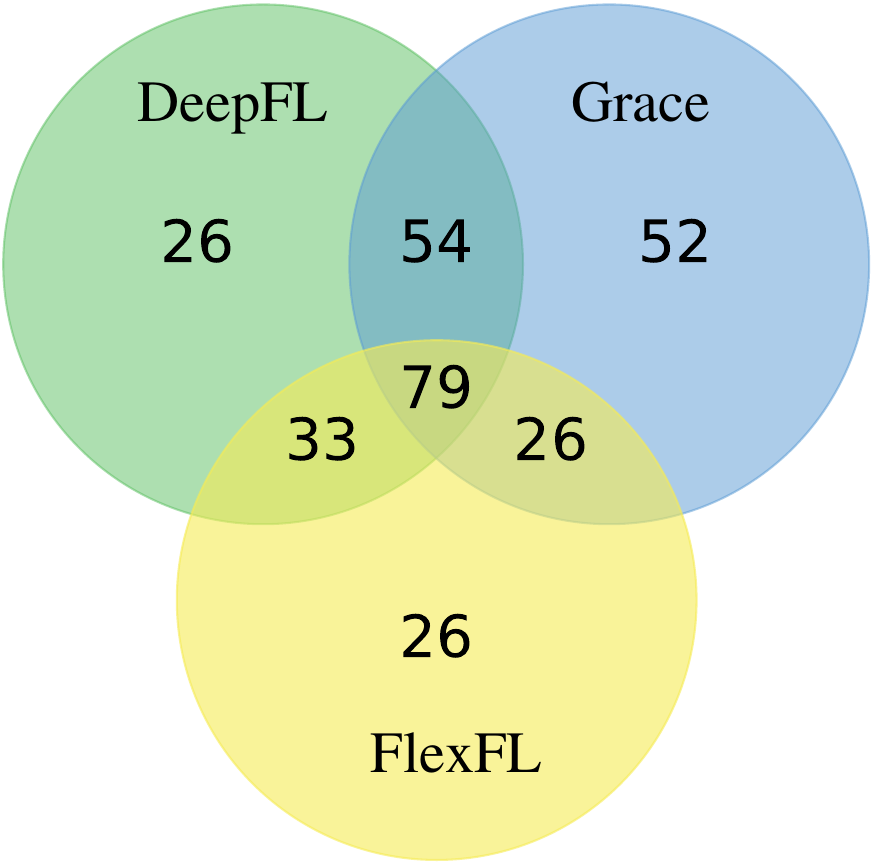}
\label{fig:venn_LBFL}
}
\caption{Comparison of \name with LBFL approaches on Defects4J (v1.0)}
\label{fig:LBFL}
\end{figure}

\new{
To enrich the analysis of \name, we compare it with two state-of-the-art learning-based fault localization (LBFL) techniques, i.e., DeepFL~\cite{DeepFL} and Grace~\cite{Grace}, on Defects4J (v1.0).
DeepFL~\cite{DeepFL} leverages RNN~\cite{RNN} and MLP~\cite{MLP} to perform fault localization based on the suspiciousness scores of SBFL and MBFL techniques, code complexity, and text similarity.
Grace~\cite{Grace} utilizes Gated Graph Neural Network (GGNN)~\cite{GGNN} to integrate graph-based coverage representation with fine-grained code structures.
As shown in Figure~\ref{fig:table_LBFL}, DeepFL and Grace demonstrate remarkable performance and outperform \name.
However, it is worth stressing that LBFL approaches require extensive datasets for training, which are usually not available or costly to collect in practice.
In contrast, FlexFL works in an unsupervised way. 
Therefore, the comparisons between LBFL approaches and FlexFL are unfair.
On the other hand, as shown in Figure~\ref{fig:venn_LBFL}, \name can localize 26 bugs at top-1 that DeepFL and Grace miss, revealing its unique superiority in localizing bugs through semantic understanding and reasoning.
The dialog-based process of FlexFL also makes it more explainable than DeepFL and Grace. 
Therefore, we believe FlexFL provides unique benefits compared to LBFL techniques and is complementary rather than competitive to them.}

\subsection{Threats to Validity}
\noindent\textbf{Internal Validity} concerns the possibility of data leakage and selection bias in the experiment.
\new{
While Defects4J has been commonly used in prior research, the fixes of its collected bugs may have been seen by the used LLM during pre-training, leading to the risk of data leakage.
However, as shown in Section~\ref{section:input_ablation}, \name performs poorly without buggy programs as input, supporting that the impact of data leakage on the effectiveness of \name on Defects4J is limited.
Moreover, \name achieves remarkable performance on the GHRB subset, which is free from data leakage, demonstrating that the effectiveness of \name is not simply due to the memorization of the used LLM.
Thus, we believe the threat of data leakage is limited.}

\noindent\textbf{External Validity} 
concerns whether the results presented would generalize. 
\new{
We evaluate our approach on Defects4J and GHRB, which are collected from medium-scale Java projects with tens of thousands of code lines to hundreds of thousands of code lines. 
We cannot claim that the evaluation results would generalize to projects in other programming languages or large-scale projects with millions of code lines, such as Linux Kernel. 
However, on the one hand, Java is one of the most popular programming languages, and our method is language-agnostic and can be adapted to other programming languages by implementing the corresponding function calls. 
On the other hand, Defects4J is widely used by prior fault localization studies, and both Defects4J and GHRB are collected from real-world high-quality projects. 
Our experimental results on GHRB, where the projects are larger than those in Defects4J, demonstrate the scalability of FlexFL to some degree. 
In addition, \name provides a general way to assist LLMs in code exploration from the buggy program, which is independent of the scale of repositories. 
Thus, it can be easily adopted to larger-scale projects. We plan to apply our approach to other programming languages and larger projects in the future.
}

\section{Conclusion and Future Work}
\label{section:conclusion}
We propose \name, a flexible and effective LLM-based FL technique that can work with lightweight open-source LLMs.
\name improves FL techniques in general scenarios utilizing a two-stage process: (1) narrowing down the search space of buggy code using state-of-the-art FL techniques of different families and (2) leveraging LLMs to delve deeper into understanding the root causes of bugs via double-checking the code snippets of the methods selected by the first stage.
To enable LLMs to leverage flexible types of bug-related information, \name designs prompt templates and LLM-based agents without assuming the existence of any specific type of bug-related information.
To make \name work with lightweight open-source LLMs, which suffer more from long input and usually have no out-of-the-box function call capabilities, we design a general manner of interaction with function calls that enable the construction of agents based on any chat model.
Evaluation results show that \name with a lightweight open-source LLM Llama3-8B can locate 42 and 63 more bugs than two state-of-the-art LLM-based FL approaches AutoFL and AgentFL that both use GPT-3.5. 
\name also demonstrates good generalization in working with different open-source LLMs and great practicality in locating bugs in the wild.
In the future, we plan to apply our approach to other datasets, more real-world scenarios, and other programming languages.

\section*{Acknowledgement}
\label{section:acknowledgement}
This research was supported by Zhejiang Provincial Natural Science Foundation of China under Grant No.LZ25F020003, and National Natural Science Foundation of China under Grant No. 62302437.

\balance
\bibliographystyle{IEEEtran}
\bibliography{IEEEabrv,paper}

\end{document}